# Laser Machined Ultrathin Microscale Platinum Thermometers on Transparent Oxide Substrates

*Letian Wang*[a], *Zeqing Jin*[a], *Dongwoo Paeng*[b], *Yoonsoo Rho*[a], *Jiangyou Long*[a], *Matthew Eliceiri*[a], *YS. Kim*[b], *and Costas P. Grigoropoulos* [a] **

(a) Laser Thermal Lab
Department of Mechanical Engineering
University of California, Berkeley
Berkeley, CA 94720-1740, USA

(b) Lam Research Corp.
4650 Cushing Pkwy, Fremont, CA 94538

**Abstract:**

Ultrathin microscale resistive thermometers are of key value to many applications. Here we have fabricated a laser machined 50 μm wide and 50 nm thick serpentine Pt thin film sensor capable of sensing temperatures up to 650 ºC over multiple heating and cooling cycles. Various materials and associated processing conditions were studied, including both sapphire and silica as transparent substrates, alumina and $TiO_2$ as adhesion layers, and lastly alumina and silicon oxide as capping layer. In-situ resistance monitoring helps verify the multi-cycle stability of the sensor and guide the optimization. 10 μm sized sensors can be laser machined but will not survive multiple heating and cooling cycles. We demonstrate that sensors with amorphous Ge thin layers can also repeatably measure temperatures up to 650 ºC. It is anticipated that this sensor can be used for fast and high spatial resolution temperature probing for laser processing applications.

## Introduction

Temperature measurement techniques come mainly in three categories[1]: electrical, optical and contact. Compared to the contact method, optical and electrical methods can provide a shorter response time. Resistive electrical sensing requires minimal instrumentation and prior knowledge of the material properties compared to optical methods. Furthermore, it is not affected by the surface conditions of the probed system, which is ideal for both the development of new thermal processes and integration with mass-produced devices.

High spatial resolution and ultrathin thin-film resistive thermometry is critical to various transient temperature probing tasks. Pt is a favorable high-temperature functional material as it is chemically inert and has a linearly temperature dependent resistivity. High spatial resolution Pt thermometers have been widely used in MEMS thermometers[2–5], heaters[6,7] and micro-reactors[8,9]. An ultrathin sensor structure contains less thermal mass, leading to shorter response time and minimum thermal interreference with the probed system. Beyond MEMS applications, laser based additive manufacturing's commercial adoption necessitates research and understanding of this transient process. The information of local temperature enables monitoring of local melting and solidifying processes thus improving the process control of laser-based manufacturing[10–12]. With laser sintering pushing printing resolution to the micron-scale[13], the sensing of local temperature becomes increasingly challenging. In addition, pulsed laser processing [14–20] requires the thickness of the sensor to be thin enough for the fast response time.

Since previous studies focus on MEMS applications on silicon[4,21], resistive sensors on transparent oxides are rarely investigated. Most biological and chemical related microfluidic chips are built on quartz and glasses, for which the thermometer applications should be addressed. More importantly, laser-based manufacturing and processing of nanomaterials happen mainly on dielectric substrates including ceramics and glasses due to their low optical absorptivity and high thermal stability. Besides these features, transparency is also of interest as it provides possibility to study the laser beam's thermal and optical effects separately by delivering the laser light from the bottom or the top of the substrate. Two typical dielectric substrates, sapphire[22] and silica were targeted for testing based on their distinctly different thermal conductivities and thermal expansion coefficients(Table.1).

Table 1 Thermal expansion coefficient for materials used in the study

| Material | Thermal Conductivity (W/mK) | Thermal Expansion ($10^{-6}$ m/(m K)) |
|---|---|---|
| Pt | 71[23] | 8.8[24] |
| Si | 130[23] | 2.6[24] |
| Fused Silica | 1.5[23] | 0.55[25] |
| Sapphire | 23.1[23] | 5.3[25] |
| Alumina | 18[23] | 8.1[25] |
| $TiO_2$ | 4.8-11.8[26] | 8.2-11.8[26] |

Femtosecond laser machining is a strong candidate for fabricating high-resolution thin film microsensors. The high-resolution sensors are conventionally fabricated with photolithography and reactive ion etching on the inert platinum, which are both time consuming and expensive. Femtosecond laser machining has been a high throughput and cost-effective industry approach for manufacturing Pt-based resistive temperature detector (RTD) elements [27]. Recent studies indicate that it is possible to machine sub-micron scale features on metal films[28]. Due to the maskless, non-vacuum, environmentally friendly, and cost-effective fabrication nature, it can be applied to large scale glass, ceramic and even flexible substrates[29]. Furthermore, on-demand laser writing offers an agile manufacturing approach, capable of placing the pattern[30], fine-tuning the resistance, and arranging multiple sensors for mapping of the temperature distribution [31].

Here, we present the successful fabrication of a stable, 50nm thick Pt high temperature sensor on transparent oxide substrates. With femtosecond laser machining, a critical dimension of 5 μm and a total size of 50 μm is achieved. Though it applies to various applications, the parameters of the current sensor are optimized towards probing laser induced transient temperature fields. The obtained 50 μm spatial size is limited by the resolution of femtosecond laser machining and consequent degradations. Nevertheless it suffices to be consistent with powder size(30-50 μm) and focused laser beam size(80 μm) in metal laser selective printing[32]. The thickness must be larger than 30nm, the experimental mean free path of electrons in Pt film[33], however 50nm thickness is required to provide enough mechanical robustness against pulsed laser induced impacts. 50nm Pt film have a thermal response time below 500ps, which prepare it for probing transient temperature induced by pulsed laser. To ensure stability upon repeatable heating, we carried out a comprehensive study on the material and associated processes for different substrates, adhesion layers, insulation layers and the existence of top layer. In-situ resistance monitoring similar to Firebaugh et al.[34] helped reveal the degradation and facilitated the optimization.

**Sensor fabrication**

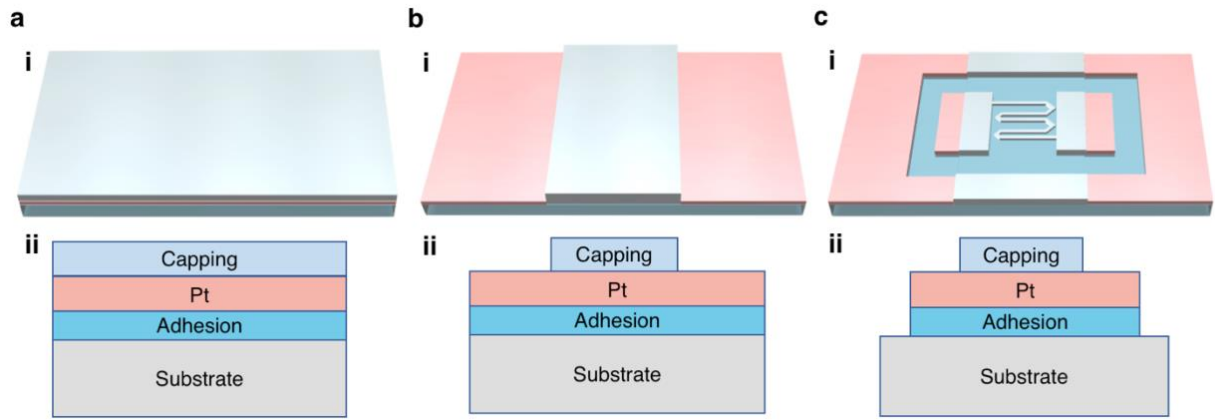

**Figure 1 Fabrication process flow for Pt thin film sensor without the top absorbing layer.** (a) Blanket deposition of adhesion, Pt and capping layers. (b) Masked etching of capping layer to expose Pt layer for electrical contact. (c) Laser micromachining of sensing pattern and electrical contact. In each panel, figure i and ii are 3D perspective and cross-sectional views respectively. "Capping" layer stands for either bare oxide capping or the case with amorphous Ge on top of oxide.

The sensor fabrication starts with blanket thin film deposition of the adhesion layer, a Pt layer and a capping layer on the selected substrates (Fig.1a). Alumina[35][36] and sapphire[37] have been reported to form epitaxial interfaces and strong adhesion with the Pt layer after annealing. For fused silica, however, adhesion to Pt is poor due to mismatched crystalline structure and thermal expansion coefficient (Table 1). Conventional Ti/ Cr metal adhesion layers suffer from oxide diffusion[38] in high-temperature annealing. $Al_2O_3$ [39,40] and $TiO_2$ [40,41] has been reported to show good adhesion with both Pt and oxide substrates without diffusion. Hence, a 7nm thick $Al_2O_3$ layer and a 15nm thick $TiO_2$ layer are deposited on a silica substrate with Atomic Layer Deposition (ALD, Cambridge) at 300 ºC for 50min. The sample is then electron-beam evaporated (CHA solutions) with 50 nm thick Pt film under $10^{-6}$ Torr. Using the tape adhesion test, we found that the Pt-sapphire combination survived multiple tape peelings, while the Pt-$Al_2O_3$-silica and Pt-$TiO_2$-silica survived one cycle and Pt-silica and Pt-oxide-Si none. From Table 2, before capping the oxide, properly adhered samples (2-4) have a similar sheet resistance of 3.45 while Pt/Oxide combinations (1 and 5) have 10% higher resistance probably due to the poor adhesion.

The deposition of the capping layer reduced Pt sheet resistance by inducing additional annealing. The capping layer is important for temperature sensing as it prevents electrical shortages, chemical diffusion or mechanical damage to the underlying Pt layer. All the samples from the above processes are capped with 45nm PECVD oxide at 350 ºC for 15min. After capping the oxide, the sheet resistances dropped significantly which we posit is due to the annealing accompanied by the deposition process[38]. The additional thermal processing will anneal the sample to ameliorate defects and increase the grain size [42]. The specific resistance of Pt thin film is as high as 19.1 μΩ·cm (before capping) and as low as 12.9 μΩ·cm (after capping). Though larger than 10.7 μΩ·cm from the bulk Pt[43], it agrees well with the results obtained on 60nm Pt film with 18.7 μΩ·cm (before) and 12.6-13.3 μΩ·cm (after annealing at 300 ºC) by Schmid et al[44]. Similar results are also obtained for 180nm Pt film with 18 μΩ·cm (before) and 14.5 μΩ·cm (after annealing 700 ºC) by Resnik et al[5]. For $TiO_2$, the resistance is not reduced significantly which could be due to the unique structural changes of $TiO_2$. [34][45] The effect of different capping layer thicknesses and processing times are investigated and discussed in the results section.)

Table 2 Sheet resistance of Pt film on different substrates before and after capping

| No | Substrates | Adhesion Layers | Before Capping | | Capping Oxide | |
|---|---|---|---|---|---|---|
| | | | Sheet Resistance (Ω/sq) | Specific resistance (μΩ·cm) | Sheet resistance (Ω/sq) | Specific resistance (μΩ·cm) |
| 1 | Silica | - | 3.81 | 19.05 | 2.58 | 12.9 |
| 2 | Silica | $TiO_2$ | 3.51 | 17.55 | 3.42 | 17.1 |
| 3 | Silica | $Al_2O_3$ | 3.46 | 17.3 | 2.67 | 13.35 |
| 4 | Sapphire | - | 3.37 | 16.85 | 2.70 | 13.5 |
| 5 | Silicon with oxide | - | 3.82 | 19.1 | 2.93 | 14.65 |

An additional Ge thin film is deposited on top of the capping layer to demonstrate the sensor's performance for applications. Germanium is an important high mobility and functional electronic material[46] and has been the subject of various laser annealing studies [47,48]. Therefore, Ge is selected as the target material and its melting point is high enough to withstand 900 ºC, which is valid for most laser processing applications. Electron beam evaporation is applied to deposit thin amorphous germanium films to avoid ion bombardments. Afterwards the sample is shadow masked in the center and etched with peroxide and buffered HF to expose part of the Pt layer for electrical contact (Fig. 1b).

After masked etching, a femtosecond pulsed laser (800nm, 1 kHz, Spitfire, Spectra-physics) is used to machine the serpentine sensor and the electrical contact pads (Fig. 1c). We eliminated the additional metal contacts to simplify the process and reduce degradation [36]. High-temperature compatible silver-paste (Ted Pella) serves as contact forming agent between the Pt contact pad and the lead wires. However, for probing laser processing, as probed resistance is the summation of both sensing unit and the contact pad, high resistance should be introduced in the sensing unit to ensure negligible error caused by contact pad. Serpentine sensor structure is designed and fabricated with a femtosecond laser under 100X objective lens (Fig.2bi). Samples are first machined to a 4mm scale, and then further shrunk to 10 μm. For the smaller, 10μm sensor, the side roughness in high-resolution machining is significant compared to the line width, which is caused by the non-uniform film quality and the stage motion stability. Further optimization can be obtained through prolonged annealing after machining.

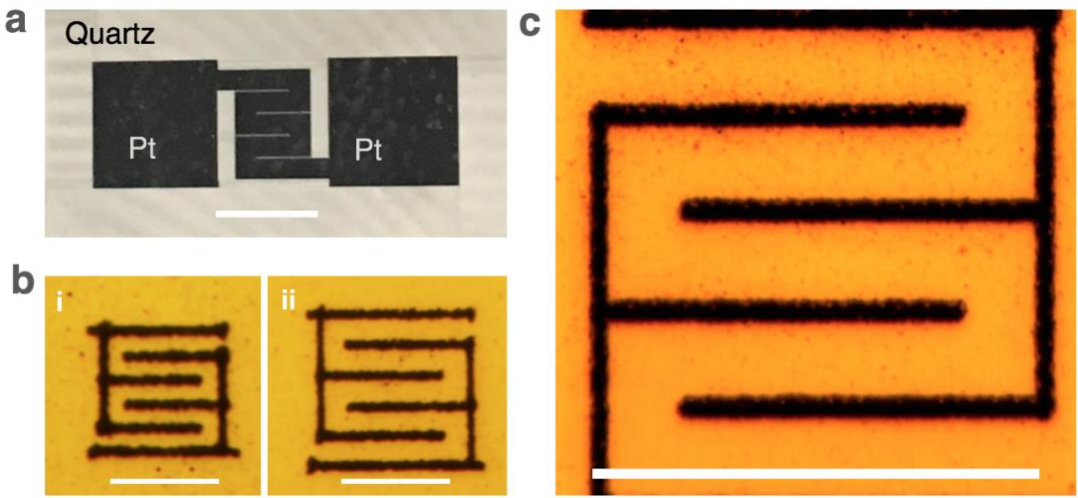

**Figure 2 Femtosecond laser machining of the Pt sensor.** (a) overview of the laser machined sample with 4mm size. (b)femtosecond machined ( i)10μm, (ii)15μm sized serpentine sensor pattern. (c) femtosecond laser machined 50μm sensor. The scan speed is 0.1mm/s. The scale bars in a-c are 5mm, 10 μm and 50 μm, respectively.

As the samples are properly insulated from oxidation, a tube furnace (Lindberg) is used to heat the sample in ambient air conditions for the calibration of Temperature Coefficient of Resistance (TCR). A separate K-type high-temperature compatible thermocouple is inserted into the tube close to the sample. The temperature and voltage change on the resistor are recorded *in-situ* through digital data logger (HIOKI LR8431). We carried out reversible heating and cooling to calibrate the TCRs of the sample using a maximum ramping rate 50C/min. The sensor's TCRs are distributed near 2500-2800 ppm/K, which agree with results of the 200nm Pt thin film[49] annealed between 250-450 °C. The *in-situ* study helps to monitor the failure and the hysteresis of the sensor under the high-temperature annealing.

**Results**

**Different Substrates' Effect on Thermal Stability**

Consistent with the adhesion tests, Pt sensors on sapphire substrates demonstrate better thermal cycling stability compared to those on silica+$TiO_2$ and silica+$Al_2O_3$. In Fig. 3, the Pt sensor on sapphire has demonstrated almost identical heating and cooling TCR in 2 cycles up to 700 °C (Fig. 3a-b). Silica+$TiO_2$ combination shows significant hysteresis effects resembling the results from Pt on the Ti adhesion layer[34]. The hysteresis is also increased after the 2$^{nd}$ heating and cooling cycle. (Fig. 3c ii)). No chemical diffusion is observed for $TiO_2$ under Pt up to 800 °C [41] and agglomeration is considered to be the main degradation mechanism, which has been proved in previous studies[34]. As it is later addressed on other substrates, a detailed investigation is not pursued. Upon further consideration of the $TiO_2$ absorptivity [50] and its semiconducting behavior [51], this material is not further pursued as the adhesion layer for Pt on silica. For silica + $Al_2O_3$, no hysteresis is seen in the cycling but significant failure is observed in the 2$^{nd}$ cycle. The failure is later addressed with longer annealing time (Fig. 4) and its cause is also discussed in the next section. Identical experiments were carried out on Si-$SiO_2$ substrates and a similar hysteresis was observed. In summary, the sapphire and silica+ $Al_2O_3$ substrates show promise for high-temperature sensing.

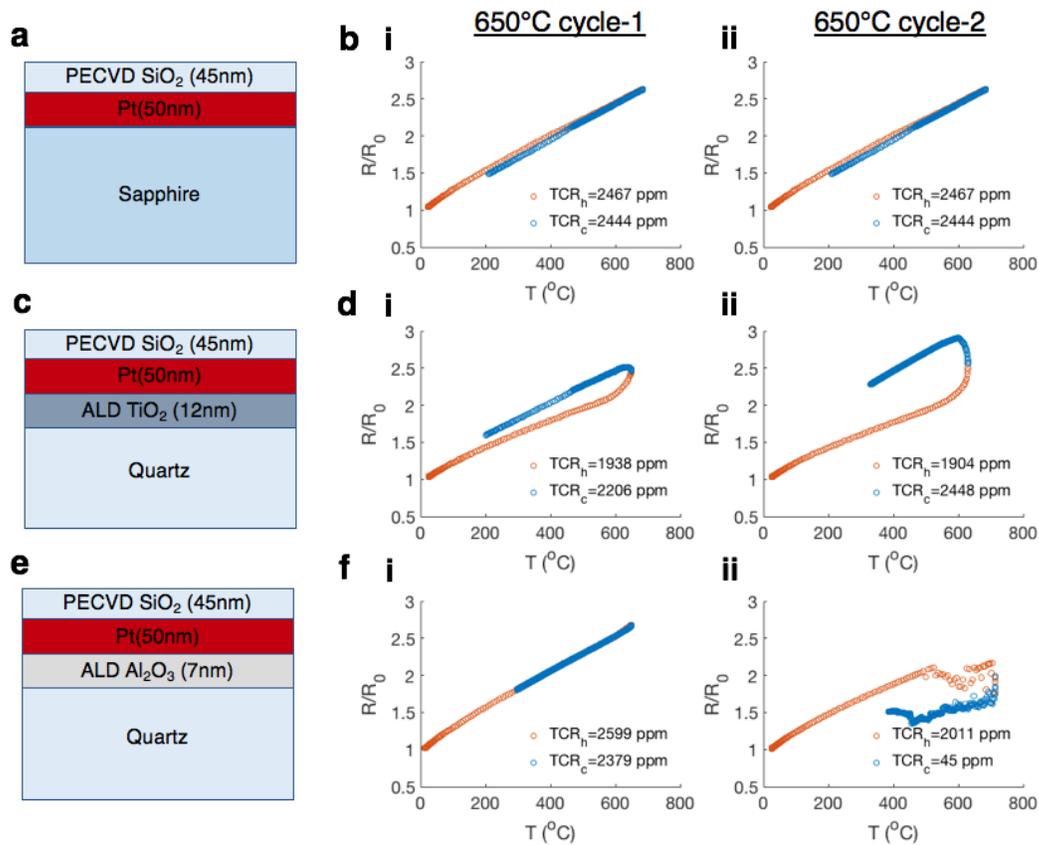

**Figure 3 Temperature dependent resistance calibration of Pt sensor on sapphire, silica+TiO$_2$ and silica+Al$_2$O$_3$ substrates**. (**a**) schematics of the sensor on a sapphire substrate and (**b**) its resistance history upon 2 cycles. (i-ii) of heating and cooling at 650 $^o$C. (**c**) Silica+TiO$_2$ substrates and (**d**) its resistance history. (**d**) Silica+Al$_2$O$_3$ substrates and (**e**) its resistance history. All the samples are capped with 45nm PECVD oxides. The heating and cooling curves are labeled with red and blue, and the TCRs are plotted in the legends.

**Capping Layer Effects on the Thermal Stability**

Low cost and low thermal conductivity make silica an important substrate for laser processing. The aforementioned initial attempt with silica + Al$_2$O$_3$ substrate, however, only survived the 1$^{st}$ cycle of heating. Based on the possible sources of degradation, two improvements are investigated, namely depositing a thick oxide or an additional thin ALD . Firstly, a thicker capping oxide layer offers stronger surface constraints to reduce agglomeration and deformation. Experimental results (Fig.4a) show 110nm capping layer exhibited reduced hysteresis over multiple annealing cycles and achieved repeatable results (Fig. 4a-b). Secondly, given the good adhesion between Pt with Al$_2$O$_3$, symmetrically placing Al$_2$O$_3$ interfacial layers on both sides of the Pt layer is anticipated to improve stability. A 7nm thick ALD Al$_2$O$_3$ layer is deposited on top of Pt thin films and repeatable results are obtained as well (Fig. 4c-d). The successful results show that neither thickness nor the choice of capping material is the critical requirement for repeatable thermal results. Alternatively, since both PECVD 110nm oxide(35min) and ALD $Al_2O_3$ (40min) involve at least two times longer thermal annealing than PECVD 45nm oxide(17min), the effective annealing time is considered as the main factor for stability. With increased number of heating cycles, the hysteresis effect is reduced, leading to less difference in TCRs for heating and cooling. Further, the overall value of TCRs increases, agreeing with the literature[49]. Insufficient annealing time means larger residual stress [52] inside Pt and between layers, which contribute to the stronger hysteresis manifested in Fig. 3fii. However, the failure of Fig. 3e-f indicates that the stress is released properly only when constant temperature annealing is implemented before any fast ramping heating. Otherwise, faster ramping and cooling does not help and may lead to sensor failure.

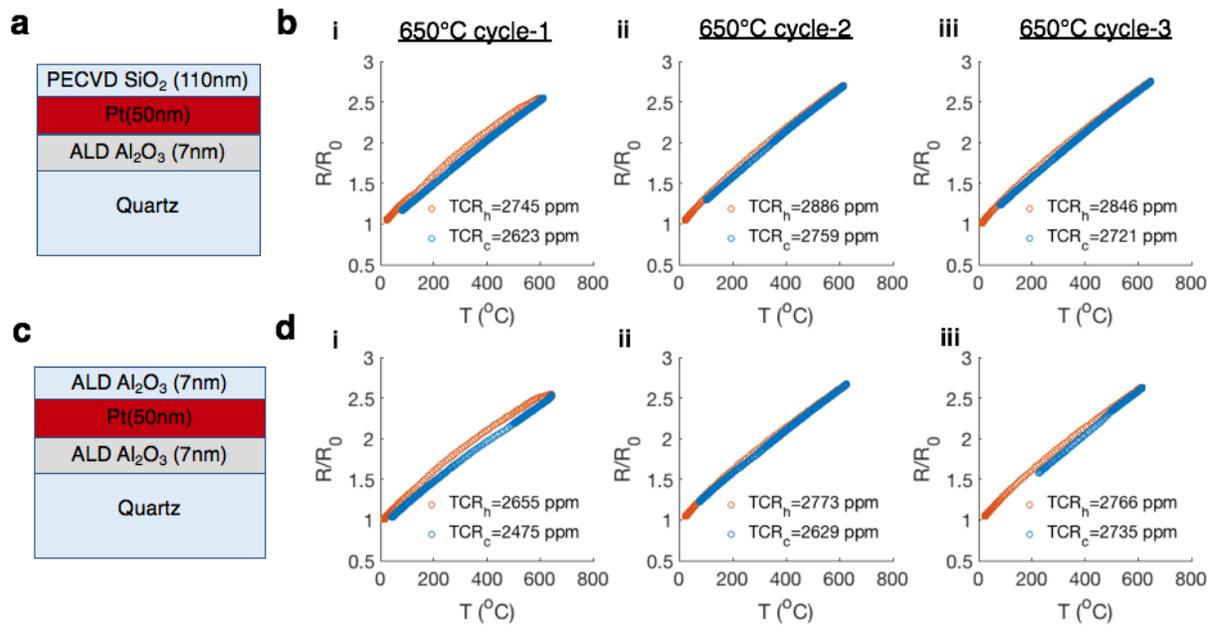

**Figure 4 Temperature dependent resistance calibration of silica+$Al_2O_3$ substrate capped with different capping layers**. (**a**) schematics of structures with PECVD 110nm $SiO_2$ as capping layer and (**b**) the resistance history upon three cycles of heating and cooling at 650 °C labeled with (i-iii). (**c**) schematics of structures with ALD 7nm $Al_2O_3$ as capping layer and (**d**) the resistance history upon the same processing as (**b**).

**High-Temperature Annealing**

Higher sensor temperature range is favorable as laser processing of semiconductor thin films can easily reach 900 °C[53,54]. We first confirm the Pt sensor on sapphire substrate survived three cycles of heating and cooling at 650 °C(Fig.5). For another identical sample, we ramped the temperature to 900 °C after one cycle of 650 °C. A significant deviation is generated at around 800 °C and similar hysteresis as silica + $TiO_2$ is observed afterward. An extreme hysteresis is generated at the 3$^{rd}$ cycle, indicating the failure of the sensor. Therefore we conclude that agglomeration is generated at high annealing temperature above 800 ºC . The sensor's maximum temperature range for steady-state measurement is between 650-800 ºC .

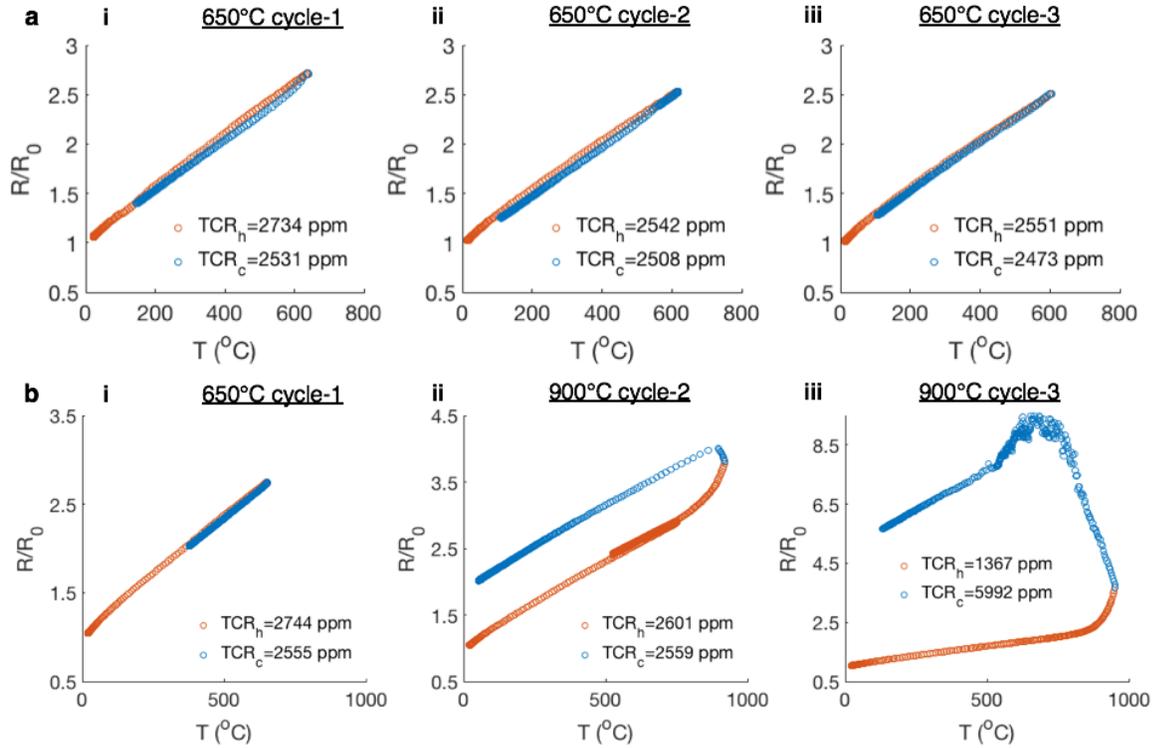

**Figure 5 Temperature dependent resistance calibration of Pt sensor on sapphire substrates with different maximum temperatures**. (**a**) 650ºC (**b**) 900 ºC with 1-3 cycles labeled with (i-iii)

**Sensor with different sizes**

Miniature sensors are ultimately required for probing focused laser-induced temperature fields. To study the minimum sensor size, we designed the sample to 200μm, 50μm and 15μm sizes. All the sensors are manufactured on silica-alumina substrates and capped with an ALD alumina thin layer. The measured TCR and drifting characteristics during repeated heating and cooling cycles are listed in Table 3. It is clear the previous annealing condition for 6mm sensor (Fig. 4c) is not enough to stabilize the 200um sensor, leading to hysteresis. Fortunately, the TCR is stabilized and the hysteresis is removed upon more heating and cooling cycles. For the 50um sensor, the TCR is not stabilized upon the 3$^{rd}$ cycle and hysteresis is generated. Smaller size induced stress concentration increased the requirement of stress release and annealing. Annealing before laser machining is shown to provide a proper solution and is discussed in the next section.

When the size shrunk down to 15μm, significant electrical shorting as well as diverging heating and cooling curves were observed, where the resistance dropped dramatically, returned to normal, and fell again afterwards. Microscopic images indicate thin-film flaking (Fig.6) is the cause of the shorting effects. Since the flake is easily detached after vibration, we note the sensor completely lost conductance after being disconnected from the measurement apparatus. The laser ablation induced shock wave delamination effects[55,56] reduced the Pt stripe's mechanical integrity and adhesion to the substrate. When the size of the sensor is reduced to 15μm, such effects will be critical and lead again to sensor failure. Similar to the above problem, annealing before laser machining and longer annealing are believed to further reduce the minimum sensor size.

Table 3 TCR measurement for the sensors with different sizes

| Size(μm) | Cycle | TCR(h/c) (ppm/K) | Hysteresis |
|---|---|---|---|
| 200 | 1 | 2403/2083 | Yes |
| 200 | 2 | 2614/2197 | No |
| 200 | 3 | 2507/2091 | No |
| 50 | 1 | 2102/2093 | No |
| 50 | 2 | 2298/2189 | No |
| 50 | 3 | 2303/1993 | Yes |
| 15 | 1 | 1703/1203 | Yes |

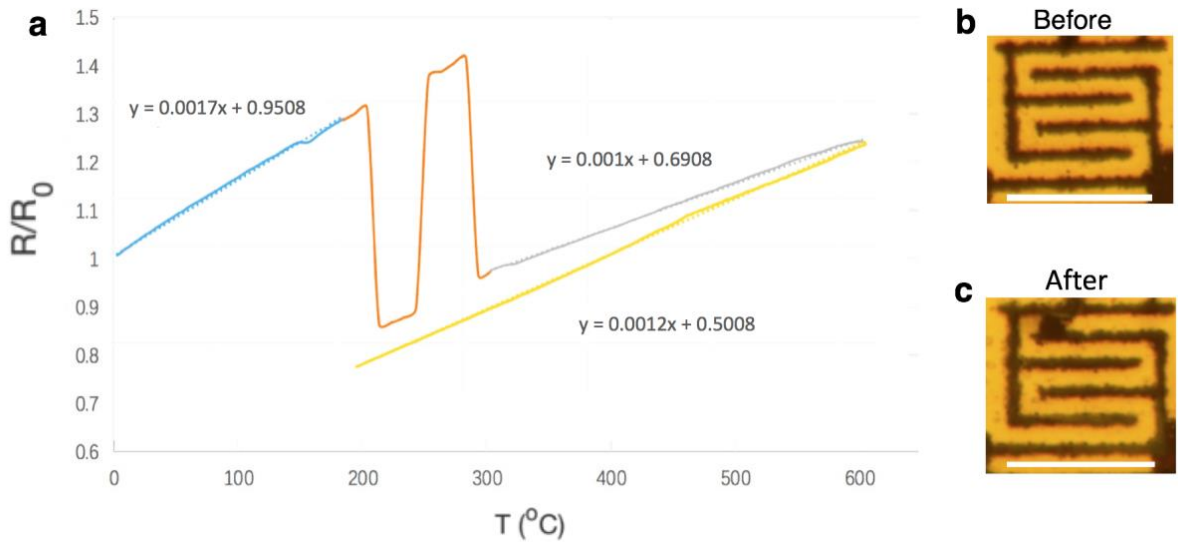

**Figure 6 Failure of 15µm sized Pt sensor under one single heating and cooling cycle**. (**a**) temperature dependent resistance measurement during 1st cycle heating and cooling. "Blue", "orange" and "grey" stand for heating process and "yellow" stands for cooling. Optical bright-field image of the sample (**b**) before 1st cycle testing and (**c**) after 1st cycle testing. The scale bars in (b-c) are 15 µm

**Microscale Sensor with Top Layer**

We investigated the thermal stability of the sensor with an amorphous Ge layer deposited on top, and the results are listed in Table 4. The original sensor is capped with PECVD 45nm thick oxide and hence no hysteresis is anticipated for the 1st cycle. However, a sensor with an amorphous Ge layer presented significant TCR hysteresis in the 1st cycle. The Ge layer introduced additional stresses during deposition and also restricted the stress release during the laser machining process. However, at the 2nd and 3rd cycle, the TCR stabilized without any hysteresis as the stress is properly released. For 50µm size sensors, the sensor obtained just slight improvement at the 2nd and 3rd cycle. We presume that the focused laser ablation generates sufficiently high stresses, and once these stresses are coupled with top layer confinement they cause irreversible degradation. As a solution, we propose annealing the sample before laser machining to release the stress and reduce the defects. A thermally stable 50µm sensor with a top layer is successfully demonstrated.

Table 4 TCR measurement for the sensors with Ge as the top layer

| Size(μm) | Cycle | TCR(h/c) (ppm/K) | Machine-Anneal sequence | Hysteresis |
|---|---|---|---|---|
| 200 | 2nd | 2510/2634 | MA | No |
| 200 | 3rd | 2602/2621 | MA | No |
| 50 | 2nd | 2198/1667 | MA | Yes |
| 50 | 3rd | 2093/2406 | MA | Yes |
| 50 | 2nd | 2513/2399 | AM | No |
| 50 | 3rd | 2507/2523 | AM | No |

**Conclusion**

We have developed fabrication processes for a stable 50nm thick microscale Pt thermometer on transparent substrates. The silica substrate with an ultrathin $TiO_2$ adhesion layer gives an increasing hysteresis response of the TCR with repeated heating and cooling. Sapphire and silica with alumina adhesion provide repeatable TCRs for cycling tests at up to 650 ºC, which fails at 900 ºC. Laser machined sensors reach a minimum size of 10μm. 15μm size thermometers will fail with irreversible film flaking. Only sensors of at least 50μm, however, survived repeated heating and cooling. Annealing before laser machining resolved the hysteresis problem induced by deposition of an additional Ge thin film and provide a strategy for further reduction of sensor size. A 50um sensor with a thin Ge layer on top shows repeatable TCR performance and is ready for measurement of laser processing.

**Acknowledgment**

The nanofabrication and SEM were carried out at the Marvell Nanofabrication Laboratory and the California Institute of Quantitative Bioscience (QB3) of UC Berkeley. The authors would like to thank Chris Zhao from in Novellution Technologies, Inc. for helpful discussion on the sample fabrication. The work received financial support from Lam Research Corp..

# Reference

[1] D.L. Blackburn, Temperature measurements of semiconductor devices - a review, Twent. Annu. IEEE Semicond. Therm. Meas. Manag. Symp. (IEEE Cat. No.04CH37545). 20 (2004) 70–80. doi:10.1109/STHERM.2004.1291304.

[2] L. Zhu, A. Fiorino, D. Thompson, R. Mittapally, E. Meyhofer, P. Reddy, Near-field photonic cooling through control of the chemical potential of photons, Nature. 566 (2019) 239.

[3] D. Thompson, L. Zhu, R. Mittapally, S. Sadat, Z. Xing, P. McArdle, M.M. Qazilbash, P. Reddy, E. Meyhofer, Hundred-fold enhancement in far-field radiative heat transfer over the blackbody limit, Nature. 561 (2018) 216.

[4] L. Cui, W. Jeong, S. Hur, M. Matt, J.C. Klöckner, F. Pauly, P. Nielaba, J.C. Cuevas, E. Meyhofer, P. Reddy, Quantized thermal transport in single-atom junctions, Science. 355 (2017) 1192 LP-1195. http://science.sciencemag.org/content/355/6330/1192.abstract.

[5] D. Resnik, D. Vrtačnik, M. Možek, B. Pečar, S. Amon, Experimental study of heat-treated thin film Ti/Pt heater and temperature sensor properties on a Si microfluidic platform, J. Micromechanics Microengineering. 21 (2011). doi:10.1088/0960-1317/21/2/025025.

[6] G. Bernhardt, C. Silvestre, N. Le Cursi, S.C. Moulzolf, D.J. Frankel, R.J. Lad, Performance of Zr and Ti adhesion layers for bonding of platinum metallization to sapphire substrates, Sensors Actuators, B Chem. 77 (2001) 368–374. doi:10.1016/S0925-4005(01)00756-0.

[7] G.-S. Chung, J.-M. Jeong, Fabrication of micro heaters on polycrystalline 3C-SiC suspended membranes for gas sensors and their characteristics, Microelectron. Eng. 87 (2010) 2348–2352.

[8] R.M. Tiggelaar, J.W. Berenschot, J.H. De Boer, R.G.P. Sanders, J.G.E. Gardeniers, R.E. Oosterbroek, A. Van Den Berg, M.C. Elwenspoek, Fabrication and characterization of high-temperature microreactors with thin film heater and sensor patterns in silicon nitride tubes, Lab Chip. 5 (2005) 326–336. doi:10.1039/b414857f.

[9] R. Srinivasan, I. Hsing, P.E. Berger, K.F. Jensen, S.L. Firebaugh, M.A. Schmidt, M.P. Harold, J.J. Lerou, J.F. Ryley, Micromachined reactors for catalytic partial oxidation reactions, AIChE J. 43 (1997) 3059–3069.

[10] M. Islam, T. Purtonen, H. Piili, A. Salminen, O. Nyrhilä, Temperature profile and imaging analysis of laser additive manufacturing of stainless steel, Phys. Procedia. 41 (2013) 835–842.

[11] D. Hu, R. Kovacevic, Modelling and measuring the thermal behaviour of the molten pool in closed-loop controlled laser-based additive manufacturing, Proc. Inst. Mech. Eng. Part B J. Eng. Manuf. 217 (2003) 441–452.

[12] B. Yuan, G.M. Guss, A.C. Wilson, S.P. Hau-Riege, P.J. DePond, S. McMains, M.J. Matthews, B. Giera, Machine-Learning-Based Monitoring of Laser Powder Bed Fusion, Adv. Mater. Technol. 3 (2018) 1800136.

[13] H. Exner, M. Horn, A. Streek, F. Ullmann, L. Hartwig, P. Regenfuß, R. Ebert, Laser micro sintering: A new method to generate metal and ceramic parts of high resolution with sub-micrometer powder, Virtual Phys. Prototyp. 3 (2008) 3–11.

[14]  B. Nie, L. Yang, H. Huang, S. Bai, P. Wan, J. Liu, Femtosecond laser additive manufacturing of iron and tungsten parts, Appl. Phys. A. 119 (2015) 1075–1080.

[15]  S. Han, S. Hong, J. Ham, J. Yeo, J. Lee, B. Kang, P. Lee, J. Kwon, S.S. Lee, M. Yang, Fast Plasmonic Laser Nanowelding for a Cu-Nanowire Percolation Network for Flexible Transparent Conductors and Stretchable Electronics, Adv. Mater. 26 (2014) 5808–5814.

[16]  D. Paeng, J. Yoo, J. Yeo, D. Lee, E. Kim, S.H. Ko, C.P. Grigoropoulos, Low-Cost Facile Fabrication of Flexible Transparent Copper Electrodes by Nanosecond Laser Ablation, Adv. Mater. 27 (2015) 2762–2767.

[17]  L. Wang, Y. Rho, W. Shou, S. Hong, K. Kato, M. Eliceiri, M. Shi, C.P. Grigoropoulos, H. Pan, C. Carraro, D. Qi, Programming Nanoparticles in Multiscale: Optically Modulated Assembly and Phase Switching of Silicon Nanoparticle Array, ACS Nano. 12 (2018) 2231–2241. doi:10.1021/acsnano.8b00198.

[18]  D. Qi, S. Tang, L. Wang, S. Dai, X. Shen, C. Wang, S. Chen, Pulse laser-induced size-controllable and symmetrical ordering of single-crystal Si islands, Nanoscale. 10 (2018) 8133–8138. doi:10.1039/C8NR00210J.

[19]  D. Qi, D. Paeng, J. Yeo, E. Kim, L. Wang, S. Chen, C.P. Grigoropoulos, Time-resolved analysis of thickness-dependent dewetting and ablation of silver films upon nanosecond laser irradiation, Appl. Phys. Lett. 108 (2016) 211602. doi:10.1063/1.4952597.

[20]  J. Long, P. Fan, M. Zhong, H. Zhang, Y. Xie, C. Lin, Superhydrophobic and colorful copper surfaces fabricated by picosecond laser induced periodic nanostructures, Appl. Surf. Sci. 311 (2014) 461–467.

[21]  Chang C.W., Okawa D., Majumder A., Zettl A., Solid-state thermal rectifier, Science. 314 (2006) 1121–1124. doi:10.1126/science.1132898.

[22]  R. Vargas, T. Goto, W. Zhang, T. Hirai, Epitaxial growth of iridium and platinum films on sapphire by metalorganic chemical vapor deposition, Appl. Phys. Lett. 65 (1994) 1094–1096. doi:10.1063/1.112108.

[23]  R.W. Powell, C.Y. Ho, P.E. Liley, Thermal conductivity of selected materials, US Department of Commerce, National Bureau of Standards Washington, DC, 1966.

[24]  D.R. Lide, CRC handbook of chemistry and physics, CRC Boca Raton, 2012.

[25]  Coefficients of Linear Thermal Expansion, Eng. ToolBox. (2003). https://www.engineeringtoolbox.com/linear-expansion-coefficients-d_95.html.

[26]  A. Materials, Titanium Dioxide-Titania(TiO2), (2018). doi:https://www.azom.com/properties.aspx?ArticleID=1179.

[27]  B. GLINIECKI, Platinum Temperature Sensor Technology, Trend Watch Mater. Asembly. (2013) T5. https://www.heraeus.com/media/media/group/doc_group/products_1/hst/usa_only/technical_articles/DesignNews.pdf.

[28]  A. Wang, L. Jiang, X. Li, Y. Liu, X. Dong, L. Qu, X. Duan, Y. Lu, Mask-Free Patterning of High-Conductivity Metal Nanowires in Open Air by Spatially Modulated Femtosecond Laser Pulses, Adv. Mater. 27 (2015) 6238–6243. doi:10.1002/adma.201503289.

[29] Y. Nakajima, K. Obata, M. Machida, A. Hohnholz, J. Koch, O. Suttmann, M. Terakawa, Femtosecond-laser-based fabrication of metal/PDMS composite microstructures for mechanical force sensing, Opt. Mater. Express. 7 (2017) 4203. doi:10.1364/OME.7.004203.

[30] Q. Chen, T. Tong, J.P. Longtin, S. Tankiewicz, S. Sampath, R.J. Gambino, Novel sensor fabrication using direct-write thermal spray and precision laser micromachnining, J. Manuf. Sci. Eng. Asme. 126 (2004) 830–836. doi:Doi 10.1115/1.1813481.

[31] S. Hong, J. Yeo, W. Manorotkul, G. Kim, J. Kwon, K. An, S.H. Ko, Low-temperature rapid fabrication of ZnO nanowire UV sensor array by laser-induced local hydrothermal growth, J. Nanomater. 2013 (2013) 2.

[32] M. Brandt, The role of lasers in additive manufacturing, Elsevier Ltd, 2016. doi:10.1016/B978-0-08-100433-3.02001-7.

[33] M.M. Da Silva, a. R. Vaz, S. a. Moshkalev, J.W. Swart, Electrical Characterization of Platinum Thin Films Deposited by Focused Ion Beam, ECS Trans. 9 (2007) 235–241. doi:10.1149/1.2766894.

[34] S.L. Firebaugh, K.F. Jensen, M.A. Schmidt, Investigation of high-temperature degradation of platinum thin films with an in situ resistance measurement apparatus, J. Microelectromechanical Syst. 7 (1998) 128–135. doi:10.1109/84.661395.

[35] J. Zhang, Y. Nagao, S. Kuwano, Y. Ito, Microstructure and temperature coefficient of resistance of platinum films, Japanese J. Appl. Physics, Part 1 Regul. Pap. Short Notes Rev. Pap. 36 (1997) 834–839. doi:10.1143/JJAP.36.834.

[36] J. Han, P. Cheng, H. Wang, C. Zhang, J. Zhang, Y. Wang, L. Duan, G. Ding, MEMS-based Pt film temperature sensor on an alumina substrate, Mater. Lett. 125 (2014) 224–226. doi:10.1016/j.matlet.2014.03.170.

[37] M. Wada, T. Maeda, S. Inoue, Single Crystal Level High Temperature Coefficient of Resistance for a Platinum Thin Film on a Sapphire Substrate by Thermal Treatment, J. Japan Inst. Met. Mater. 76 (2012) 359–363. doi:10.2320/jinstmet.76.359.

[38] A. Ababneh, A.N. Al-Omari, A.M.K. Dagamseh, M. Tantawi, C. Pauly, F. Mücklich, D. Feili, H. Seidel, Electrical and morphological characterization of platinum thin-films with various adhesion layers for high temperature applications, Microsyst. Technol. 23 (2017) 703–709. doi:10.1007/s00542-015-2715-0.

[39] V. Guarnieri, L. Biazi, R. Marchiori, A. Lago, Platinum metallization for MEMS application. Focus on coating adhesion for biomedical applications, Biomatter. 4 (2014) e28822. doi:10.4161/biom.28822.

[40] S. Halder, T. Schneller, R. Waser, Enhanced stability of platinized silicon substrates using an unconventional adhesion layer deposited by CSD for high temperature dielectric thin film deposition, Appl. Phys. A Mater. Sci. Process. 87 (2007) 705–708. doi:10.1007/s00339-007-3866-3.

[41] A. Ababneh, A.N. Al-Omari, M. Marschibois, D. Feili, H. Seidel, Investigations on the high temperature compatibility of various adhesion layers for platinum, Proc. SPIE. 8763 (2013) 87631Z. doi:10.1117/12.2017333.

[42] W. Sripumkhai, S. Porntheeraphat, B. Saekow, W. Bunjongpru, S. Rahong, J. Nukeaw, Effect of

Annealing Temperature on Platinum Thin Films Prepared by Electron Beam Evaporation, J. Microsc. Soc. Thail. 24 (2010) 51–54.

[43] G. Fischer, H. Hoffmann, J. Vancea, Mean free path and density of conductance electrons in platinum determined by the size effect in extremely thin films, Phys. Rev. B. 22 (1980) 6065.

[44] U. Schmid, H. Seidel, Influence of thermal annealing on the resistivity of titanium/platinum thin films, J. Vac. Sci. Technol. A Vacuum, Surfaces, Film. 24 (2006) 2139–2146. doi:10.1116/1.2359739.

[45] F.N. Mohamed, M.S. A Rahim, N. Nayan, M.K. Ahmad, M.Z. Sahdan, J. Lias, Influence of TiO2 thin film annealing temperature on electrical properties synthesized by CVD technique, (2015).

[46] J. Li, S. Chen, D. Qi, W. Huang, C. Li, H. Lai, Energy band design for p-type tensile strained Si/SiGe multi-quantum well infrared photodetector, Optoelectron. Lett. 7 (2011) 175–177.

[47] S. Heo, S. Baek, D. Lee, M. Hasan, H. Jung, J. Lee, H. Hwang, Sub-15 nm n+/ p-Germanium Shallow Junction Formed by PH3 Plasma Doping and Excimer Laser Annealing, Electrochem. Solid-State Lett. 9 (2006) G136–G137.

[48] D. Qi, H. Liu, W. Gao, S. Chen, C. Li, H. Lai, W. Huang, J. Li, Investigations of morphology and formation mechanism of laser-induced annular/droplet-like structures on SiGe film, Opt. Express. 21 (2013) 9923–9930. doi:10.1364/OE.21.009923.

[49] R.M. Tiggelaar, R.G.P. Sanders, A.W. Groenland, J.G.E. Gardeniers, Stability of thin platinum films implemented in high-temperature microdevices, Sensors Actuators, A Phys. 152 (2009) 39–47. doi:10.1016/j.sna.2009.03.017.

[50] V.N. Kuznetsov, N. Serpone, Visible light absorption by various titanium dioxide specimens, J. Phys. Chem. B. 110 (2006) 25203–25209. doi:10.1021/jp064253b.

[51] A. Yildiz, S.B. Lisesivdin, M. Kasap, D. Mardare, Electrical properties of TiO2 thin films, J. Non. Cryst. Solids. 354 (2008) 4944–4947. doi:10.1016/j.jnoncrysol.2008.07.009.

[52] W.A. Strifler, C.W. Bates, Stress in evaporated films used in GaAs processing, J. Mater. Res. (1991).

[53] M. Hatano, S. Moon, M. Lee, K. Suzuki, C.P. Grigoropoulos, In situ and ex situ diagnostics on melting and resolidification dynamics of amorphous and polycrystalline silicon thin films during excimer laser annealing, J. Non. Cryst. Solids. 266–269 (2000) 654–658. doi:10.1016/S0022-3093(99)00768-1.

[54] D.P. Brunco, J. a Kittl, M. Thompson, Time-resolved temperature measurements using thin film metal thermometers, Rev. Sci. Instrum. 64 (1993) 2615–2623.

[55] S. Haas, G. Schöpe, C. Zahren, H. Stiebig, Analysis of the laser ablation processes for thin-film silicon solar cells, Appl. Phys. A Mater. Sci. Process. 92 (2008) 755–759. doi:10.1007/s00339-008-4560-9.

[56] P. Lorenz, T. Smausz, T. Csizmadia, M. Ehrhardt, K. Zimmer, B. Hopp, Shadowgraph studies of laser-assisted non-thermal structuring of thin layers on flexible substrates by shock-wave-induced delamination processes, Appl. Surf. Sci. 336 (2015) 43–47. doi:10.1016/j.apsusc.2014.09.114.